\documentclass[prd,aps,showpacs,preprintnumbers,amssymb]{revtex4}
\usepackage{graphicx}
\usepackage{dcolumn}
\usepackage{bm}
\begin{document}
\title{
Leptonic CP violation in a two parameter model
}
 \author{Samina S. {\sc Masood}}
 \email{masood@oswego.edu}
 \affiliation{Department of Earth Sciences, SUNY Oswego,
Oswego, NY 13126}                                           
 \author{Salah {\sc Nasri}}
 \email{snasri@physics.umd.edu}
 \affiliation{Department of Physics, University of Maryland, College Park, MD
 20742-4111,}
 \author{Joseph {\sc Schechter}}\email{schechte@phy.syr.edu}
\affiliation{Department of Physics, Syracuse University,
Syracuse, NY 13244-1130}
\date{December 2004}

\begin{abstract}
We further study the ``complementary" Ansatz, Tr$(M_\nu)$=0, for a
prediagonal light Majorana type neutrino mass matrix. Previously, this
was studied for the CP conserving case and the case where the two
Majorana type CP violating phases were present but the Dirac type
CP violating phase was neglected. Here we employ a simple geometric
algorithm which enables us to ``solve" the Ansatz including all three
CP violating phases. Specifically, given the known neutrino oscillation data
and an assumed two parameter (the third neutrino mass $m_3$
and the Dirac CP phase $\delta$) family of inputs we predict
the neutrino masses and Majorana CP phases. Despite the two parameter
ambiguity, interesting statements emerge.
 There is a characteristic pattern of interconnected masses and CP phases.
For large $m_3$ the three neutrinos are approximately degenerate. The only
possibility for a mass hierarchy is to have $m_3$ smaller than the other two.
A hierarchy with $m_3$ largest is not allowed.
 Small CP violation is possible only near two special values of $m_3$. Also, the 
neutrinoless double beta decay parameter is approximately bounded as
 0.020 eV $<|m_{ee}|<$ 0.185 eV. As a byproduct of looking at physical amplitudes
we discuss an alternative parameterization of the lepton mixing
 matrix which results in simpler formulas. The physical meaning of this
parameterization is explained.
\end{abstract}
\pacs{14.60 Pq, 11.30 Er, 13.15 +g}
\maketitle

\section{Introduction}

The remarkable experimental achievements (for some recent examples
see refs. \cite{{kamland},{sno},{k2k}})
 relating
to neutrino oscillations \cite{nos} have brought much closer to reality
the goal of determining the ``light" neutrino masses
and the presumed 3 $\times$ 3 lepton mixing matrix.
It is possible that more than three light neutrinos
are required in order to understand the results of the LSND
experiment \cite{lsnd}. However, we consider it reasonable,
before deciding on this,
to wait for further
supporting evidence as should be supplied soon by the
MiniBooNE collaboration \cite{mbc}. The mixing matrix contains three
mixing angles and, if the neutrinos are considered to be Dirac
type fermions, a single CP violation phase. That would be completely 
analogous to the situation prevailing in the quark sector of the
electroweak theory. But it seems very interesting to consider
the possibility that the three light neutrinos are Majorana
type fermions. This involves only half as many fermionic degrees of
freedom and would be mandated if neutrinoless double beta decay
were to be conclusively established. The Majorana neutrino
scenario implies the existence of two additional CP violation 
phases\cite{{sv}, {bhp}, {dknot}, {svandgkm}}. 
Then nine quantities (beyond the charged lepton masses)
would be required for the specification of the lepton sector:
three neutrino masses, three mixing angles and three CP violation
phases.

 According to a recent
analysis \cite{mstv} it is possible to extract from the data to good accuracy, 
two squared neutrino mass differences: $m_2^2-m_1^2$ and $|m_3^2-m_2^2|$,
and two inter-generational mixing angle squared sines: $s_{12}^2$
 and $s_{23}^2$. 
Furthermore the inter-generational mixing parameter $s_{13}^2$ is found to
be very small. Thus 5 out of 9 quantities needed to describe the leptonic 
sector in the Majorana neutrino scenario can be considered as ``known".
For many purposes it is desirable to get an idea of the remaining
4 parameters. As an aid in partially determining the 
other parameters, a so-called ``complementary Ansatz" was proposed
\cite{bfns, hz, ro, nsm}. The name arises from the fact that if CP
violation is neglected, the Ansatz determines (up to two
different cases) all three neutrino masses,
given the two known squared mass differences.

This complementary Ansatz simply reads,
\begin{equation}
Tr(M_\nu) = 0.
\label{ansatz}
\end{equation}
Here $M_\nu$ is the symmetric, but in general complex,
prediagonal Majorana neutrino mass matrix. It is brought to
diagonal form by the transformation,
\begin{equation}
U^TM_{\nu}U=diag(m_1,m_2,m_3)= {\hat M_\nu},
\label{diagonalize}
\end{equation}
where $U$ is a unitary matrix and the $m_i$ may be chosen as real, 
positive.
We impose the condition in a basis where the charged leptons are diagonal
so that U gets identified with the lepton mixing matrix.
 
Since Eq. (\ref{ansatz}) comprises two real equations, it gives two conditions
on the unknown 4 parameters of the lepton sector. In other words, the lepton
sector is described by a two parameter family of solutions. This can be 
approximately  simplified \cite{bfns, ro, nsm}   by noting that
the effects of one CP violation phase get
suppressed when the small quantity $s_{13}$ vanishes.
Then there is a one parameter family of solutions describing
the lepton sector and it is straightforward to compute physical quantities
for parameter choices which span this family. The main purpose of the 
present paper is
 to find the general solutions
 of Eq. (\ref{ansatz})
without making this approximation. This gives a two parameter family which 
allows one to study the interplay of all three CP violation phases. 

A plausibility argument supporting the complementary Ansatz is concisely
 presented in sections 2 and 3 of \cite{nsm2}. It is based on using the
SO(10) grand unification group in the approximation that the non-seesaw
neutrino mass
term dominates. The Higgs fields which can contribute to fermion masses
 at tree level are the {\bf 10}, {\bf 120} and the {\bf 126}.
 If only a single
{\bf 126} appears (but any number of the others) one has the relation
\begin{equation}
Tr(M^D-rM^E) \propto Tr(M_\nu),
\label{so10}
\end{equation} 
where $M^D$ and $M^E$ are, respectively, the prediagonal
mass matrices of the charge -1/3 quarks and charge -1 leptons, 
while $r\approx$ 3 takes account of running masses from the
 grand unified scale to about 1 GeV. Now one of the major surprises
generated by the neutrino oscillation experiments is that,
unlike the quark mixing matrix which has the form diag (1,1,1) +
O$(\epsilon)$, the lepton mixing matrix is not at all close to the unit 
matrix. This suggests a further approximation in which one takes
 $M^D$ and $M^E$ to be diagonal but allows the neutrino mass
matrix to be far from the unit matrix. Then the left hand side
of Eq. (\ref{so10}) is approximately $(m_b-3m_\tau)$, which 
is in turn close to zero. 

As we will see, the model makes a number of characteristic
predictions for the neutrino mass spectrum which should enable it to be
readily tested in the near future. A very recent review of many other 
models is given in ref. \cite{metal}.     

    For convenience, our notation (essentially the standard one) for
the lepton mixing matrix and the corresponding parameterized Ansatz is 
given in section II.

 In section III the Ansatz is solved in the sense
of providing a geometrical algorithm which, given the two input
quantities $m_3$ (third neutrino mass, taken positive) and 
$\delta$ (conventional CP violation phase in the lepton mixing matrix),
enables one to predict the other two neutrino masses as well as the other
two CP violation phases. Of course, the experimental knowledge on
the neutrino squared mass differences and CP conserving intergenerational
mixing angles are taken to be ``known". We separate the solutions into
two types I and II, depending respectively  on whether $m_3$ is the 
largest or the smallest of the neutrino masses. In addition, there is a 
discrete ambiguity corresponding to reflecting a triangle involved
in the algorithm. A ``panoramic" view of the predictions as functions
of $m_3$ and $\delta$ are presented in a convenient tabular form.
The greatest allowed value of $m_3$ is determined by a cosmology
bound. As $m_3$ decreases, a point is reached at which the type I 
solutions no longer exist. As $m_3$ decreases even further, the type
II solutions also cease to exist. The corresponding values of $m_3$ at
which these solutions become disallowed depend on the assumed value of the 
input $\delta$. This correlation is studied analytically.

    Some physical considerations are discussed in section IV.
First, the dependence on the experimentally bounded 
squared mixing angle, $s_{13}^2$ is investigated. We present also a 
chart showing the dependence of the neutrinoless double beta decay 
parameter $|m_{ee}|$ on the input parameters $m_3$ and $\delta$.
Even though the inputs are varying over a fairly large range, the
 rather restrictive approximate bounded range
 0.020 eV $<|m_{ee}|<$ 0.185 eV 
emerges from the Ansatz.

    After calculating observable quantities in the model one observes
that they depend more simply on certain linear combinations of the 
conventional ``Dirac" and ``Majorana"  CP violation phases. In section
V we discuss an alternative parameterization of the lepton mixing matrix 
in which these combinations occur directly. In this parameterization
the three phases just correspond to the three possible intergenerational 
mixings. An ``invariant" combination of these three corresponds to 
the usual ``Dirac" phase $\delta$. 

    We conclude in section VI which contains a brief summary and 
a discussion of results which emphasize some unique features of the 
present work.

\section{Parameterized complementary Ansatz}

We define the lepton mixing matrix, $K$ from the charged gauge boson 
interaction term
in the leptonic sector of the electroweak Lagrangian:
\begin{equation}
{\cal L}= \frac{ig}{\sqrt 2}W_{\mu}^-{\bar e_L}\gamma_{\mu}K\nu + H.c..
\label{weakinteraction}
\end{equation}
Note that the ``mass diagonal" neutrino fields, $\nu_i$ are related to
the fields, $\rho_i$ in the prediagonal mass basis by the
matrix equation,
\begin{equation}
\rho=U\nu.
\label{defineU}
\end{equation}                
We adopt essentially what seems to be the most common parameterization:
\begin{equation}
K=K_{exp}\omega_0^{-1}(\tau),
\label{fullmixingmatrix}
\end{equation}            
where a unimodular diagonal matrix of phases is defined as,
\begin{eqnarray}
\omega_0(\tau)&=&diag(e^{i\tau_1}, e^{i\tau_2}, e^{i\tau_3}), \nonumber \\
\tau_1 +&\tau_2& + \tau_3 = 0.
\label{majoranaphases}
\end{eqnarray}                                            
The remaining factor, $K_{exp}$ which is the only part needed for 
describing ordinary neutrino oscillations is written as the product of 
three successive two dimensional unitary transformations,
\begin{equation}
K_{exp}=\omega_{23}(\theta_{23},0)\omega_{13}(\theta_{13},-\delta)\omega_{12}(\theta_{12},0),
\label{threeomegas}
\end{equation}                                                
with three mixing angles and the CP violation phase $\delta$.
 For example in the (12) subspace
one has:
\begin{equation}
\omega_{12}(\theta_{12},\phi_{12})=\left[ \begin{array}{c c c}
cos \theta_{12}&e^{i\phi_{12}}sin \theta_{12}&0\\
-e^{-i\phi_{12}}sin \theta_{12}&cos \theta_{12}&0\\
0&0&1
\end{array} \right]
\label{onetwotransf}
\end{equation}
with clear generalization to the (23) and (13) transformations.          
Multiplying
out yields:
\begin{equation}
K_{exp}=\left[ \begin{array}{c c c}
c_{12}c_{13}&s_{12}c_{13}&s_{13}e^{-i\delta}\\
-s_{12}c_{23}-c_{12}s_{13}s_{23}e^{i\delta}&c_{12}c_{23}-s_{12}s_{13}s_{23}
e^{i\delta}&c_{13}s_{23}\\
s_{12}s_{23}-c_{12}s_{13}c_{23}e^{i\delta}&-c_{12}s_{23}-s_{12}s_{13}c_{23}
e^{i\delta}&c_{13}c_{23}\\
\end{array} \right]
\label{usualconvention}
\end{equation}                                                             
where $s_{ij} = sin \theta_{ij}\; and \; c_{ij} = cos
\theta_{ij}$.                  

Identifying $K$ with $U$ in Eq. (\ref{diagonalize}), the
Ansatz of Eq. (\ref{ansatz}) now reads,
\begin{equation}
Tr({\hat M_\nu}K_{exp}^{-1}K_{exp}^*\omega_0(2\tau))=0,
\label{ansatzwithKexp}
\end{equation}
where Eqs. (\ref{diagonalize}) and (\ref{fullmixingmatrix}) were
used. With the parameterized mixing matrix of Eq. (\ref{usualconvention})
the Ansatz finally becomes:

\begin{eqnarray}
m_1e^{2i\tau_1} \left[ 1 -2i(c_{12}s_{13})^2sin{\delta}e^{-i\delta} 
\right]
 +\nonumber\\
m_2e^{2i\tau_2} \left[ 1 -2i(s_{12}s_{13})^2sin{\delta}e^{-i\delta} 
\right]
\nonumber +\\
m_3e^{2i\tau_3} \left[ 1 +2i(s_{13})^2sin{\delta}e^{i\delta} \right] = 0.
\label{paramansatz}
\end{eqnarray}
In this equation we can choose the diagonal masses $m_1, m_2, m_3$ to be
 real positive.                                                             
Notice that setting the mixing parameter $s_{13}$ to zero eliminates the
dependence on the CP violation phase $\delta$. Then Eq. 
(\ref{paramansatz})
goes over to the simpler form studied previously \cite{nsm}.

\section{Solving the Ansatz in the general case}

   For definiteness we will use the following best fit values
for the differences of squared neutrino masses obtained in ref.
\cite{mstv}:
\begin{eqnarray}
 A \equiv m_2^2-m_1^2 &=& 6.9 \times 10^{-5} eV^2, \nonumber \\
 B \equiv |m_3^2-m_2^2| &=& 2.6 \times 10^{-3} eV^2.
\label{massdifferences}
\end{eqnarray}             
The uncertainty in these determinations is roughly $25 \%$.                              
Similarly for definiteness  we will adopt the best fit values for 
$s_{12}^2$
and $s_{23}^2$ obtained in the same analysis:
\begin{equation}
s_{12}^2 = 0.30,\quad  s_{23}^2 = 0.50.
\label{exptmixingangles}
\end{equation}                                             
 These mixing angles also have about a  $25 \%$ uncertainty.
The experimental status of $s_{13}^2$ is less accurately known. At present
only the  3 $\sigma$ bound,
\begin{equation}
s_{13}^2 \leq 0.047,
\label{onethreebound}
\end{equation}
is available. For our discussion we will consider $s_{13}^2$
to be known at a ``typical" value satisfying this bound 
and examine the sensitivity
to changing it. Of course, the experimental
determination of $s_{13}$ is a topic of great current interest.

    Previously\cite{nsm}, the (positive) mass of the third neutrino, 
$m_3$ was considered to be the free parameter. It was varied to
obtain, via the simplified Ansatz equation,
  a ``panoramic view" of the 
two independent Majorana phases (say $\tau_1$ and $\tau_2$). In the 
present case, we shall not neglect the CP violation phase $\delta$
and consider it too as a free parameter to be varied. It is necessary
to specify a suitable algorithm to treat the full Ansatz. Previously
it was noted that the simplified Ansatz could be pictured as a vector
triangle in the complex plane having  sides equal to 
corresponding neutrino
masses (See Fig. 1 of \cite{nsm}). The three internal angles were
found by trigonometry and related to the three angles made by 
the sides with respect to the positive real axis. Those in turn were
twice the three (constrained) Majorana phases. The orientation
of the triangle got determined (up to a reflection) by the constraint
in Eq. (\ref{majoranaphases}). In the present case we will also
rewrite the Ansatz equation as a vector triangle in the complex
plane. However the sides will differ from the neutrino masses.
In addition the angles will differ from twice the constrained 
Majorana phases.

    To start, we choose a value for $m_3$ and a value for the phase $\delta$.
Then we can obtain from Eqs.(\ref{massdifferences}) two different 
solutions for the other
masses $m_1$ and $m_2$. We call the solution where $m_3$ is the largest neutrino
mass, the type I case. The case where $m_3$ is the smallest neutrino mass is
designated type II. $m_1$ will be determined from the assumed value of
$m_3$ as,
\begin{equation}
m_1^2=m_3^2-A \mp B,
\label{findmone}
\end{equation}
where the upper and lower sign choices respectively refer 
to the type I and type II cases. In either case 
we find $m_2$ as,
\begin{equation}
m_2^2=A+m_1^2.
\label{findmtwo}
\end{equation}.

\begin{figure}[htbp]
\centering
{\includegraphics[width=6.5cm,height=5.5cm,clip=true]{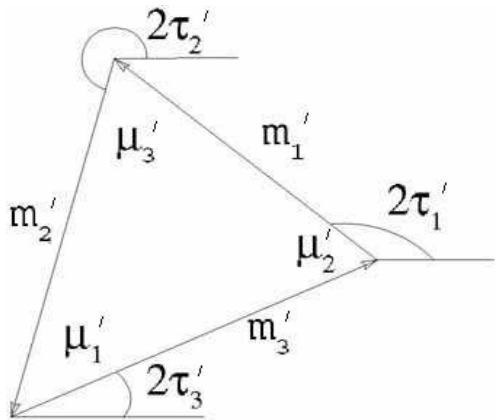}}
\caption[]{Vector triangle representing Eq. (\ref{conciseansatz}).}
\label{fig:1}
\end{figure}

Next, we redefine variables so that each of the three terms in the Ansatz
equation, (\ref{paramansatz}) is characterized by a single magnitude,
$m_i^{\prime}$ and a single phase, $2\tau_i^{\prime}$. Eq. 
(\ref{paramansatz})
then reads,   
\begin{equation}
m_1^{\prime}e^{2i\tau_1^{\prime}}+m_2^{\prime}e^{2i\tau_2^{\prime}}+
m_3^{\prime}e^{2i\tau_3^{\prime}}=0.   
\label{conciseansatz}
\end{equation}
This equation evidently represents a vector triangle in the complex plane,
as illustrated in Fig. \ref{fig:1}.
However the lengths are not the physical neutrino masses and the phases
are not twice the physical Majorana CP violation phases. The auxiliary, primed, 
masses are seen to be related to the (now known) physical masses by,
\begin{equation}
m_i^{\prime}=G_im_i,
\label{definemprime}
\end{equation}
where,
\begin{eqnarray}
G_1&=&\left[ 1-4\left( c_{12}s_{13}\right) ^{2}\sin ^{2}\delta
+4\left( c_{12}s_{13}\right) ^{4}\sin ^{2}\delta \right]^{1/2}, \nonumber \\
G_2&=&\left[ 1-4\left( s_{12}s_{13}\right) ^{2}\sin ^{2}\delta
+4\left( s_{12}s_{13}\right) ^{4}\sin ^{2}\delta \right]^{1/2}, \nonumber \\
G_3&=&\left[ 1-4s_{13}^{2}\sin ^{2}\delta +4s_{13}^{4}\sin
^{2}\delta\right]^{1/2}.
\label{defineGi}
\end{eqnarray}
Notice that, since $\delta$ has already been specified, the relations between
the $m_i^{\prime}$ and the $m_i$ are now known.
Similarly, the physical phases are related to the primed ones by,
\begin{equation}
\tau_i=\tau_i^{\prime}+F_i,
\label{findtaui}
\end{equation}
where,
\begin{eqnarray}
F_1&=&\frac{1}{2}\arctan \frac{\left( c_{12}s_{13}\right) 
^{2}\sin(2
\delta)}{1-2\left( c_{12}s_{13}\right) ^{2}\sin ^{2}\delta },
\nonumber \\
F_2&=&\frac{1}{2}\arctan \frac{\left( s_{12}s_{13}\right) ^{2}\sin(2
\delta)}{1-2\left( s_{12}s_{13}\right) ^{2}\sin ^{2}\delta },
\nonumber \\
F_3&=&-\frac{1}{2}\arctan \frac{s_{13}^{2}\sin(2\delta)}{
1+2s_{13}^{2}\sin ^{2}\delta }.
\label{defineFi}
\end{eqnarray}
Again, note that, since $\delta$ has been specified, the relations
between the $\tau_i$ and the $\tau_i^{\prime}$ are now known. Referring
to Fig. 1 we can determine the internal angles, $\mu_i^{\prime}$
by using the law of cosines. For example,
\begin{equation}
cos\mu_1^{\prime}=\frac{-(m_1^{\prime})^2+(m_2^{\prime})^2+(m_3^{\prime})^2}
{2m_2^{\prime}m_3^{\prime}}.
\label{lawofcos}
\end{equation}
Next, the auxiliary phases $\tau_i^{\prime}$ 
can be related to the internal angles just obtained as:
\begin{eqnarray}
\tau_1^{\prime}&=& \frac{1}{6} (\pi-\mu_1^{\prime}-2\mu_2^{\prime})
+ \rho , \nonumber\\
\tau_2^{\prime}&=& \frac{1}{6} (\pi+2\mu_1^{\prime}+\mu_2^{\prime})
+ \rho, \nonumber\\
\tau_3^{\prime}&=& \frac{1}{6} (-2\pi-\mu_1^{\prime}+\mu_2^{\prime})
+ \rho.
\label{thethreetauprimes}
\end{eqnarray}                                                    
The remaining still unknown parameter here is   
$\rho$ which we added to the right hand side of each 
equation. It represents the effect of an arbitrary rotation
of the whole triangle, which should not be determinable
from the internal angles. It can be determined, however,
by making use of the constraint on the physical phases,
$\sum_{i}\tau_i=0$. Notice that there is no corresponding
constraint for $\sum_{i}\tau_i^{\prime}$.
 Using Eq. 
(\ref{findtaui})
 we get,
\begin{equation}
\rho=-\frac{1}{3}\sum_{i}F_i,
\label{findbeta}
\end{equation}
where the $F_i$ are to be read from Eqs. (\ref{defineFi}).
Now the masses $m_1, m_2$ and the phases $\tau_1, \tau_2$ have been determined
by a simple algorithm, upon specification of $m_3$ and $\delta$.

As remarked above, the dependence on the CP violation phase $\delta$ is 
suppressed in the limit that the mixing parameter $s_{13}^2$ vanishes.
Hence, to illustrate this new feature, we will consider a value $s_{13}^2$
=0.04, close to the  3$\sigma$ upper bound of 0.047 \cite{mstv}.
 The predictions of 
the  neutrino masses $(m_1, m_2)$  and two independent phases
$(\tau_1, \tau_2)$, from the Ansatz for various assumed values of
$m_3$ and $\delta$ are given in Table \ref{trianglesampler}.
Representative values of $\delta$ were chosen to lie between 0
and $\pi$ since it may be observed from Eqs. (\ref{defineGi})
 and (\ref{defineFi})
that the solutions will have a periodicity of $\pi$ with respect to 
$\delta$.
   Just from the
Ansatz there is no upper bound on the value of $m_3$. However
there is a recent cosmological bound \cite{cosmobound} which
requires,
\begin{equation}
|m_1|+|m_2|+|m_3|<0.7 \textrm{ eV}. \label{cosmobound}
\end{equation}
Thus values of $m_3$ greater than about
0.3 eV are physically disfavored.
 Table \ref{trianglesampler} shows that at this value both
type I and type II solutions exist. This is true also for higher values 
of $m_3$.  The picture remains very
similar down to around $m_3=0.1$ \textrm{ eV} but as one gets
closer to roughly $0.06$ \textrm{ eV}, 
 there is a marked change.  If one further lowers $m_3$, it is found that
the type I solution no longer exists. On the other hand the type II
solution persists and does not change much until $m_3$
approaches the neighborhood of 0.001 eV. There
are no solutions for $m_3$ below this region.               
       
\begin{table}[htbp]
\begin{center}
\begin{tabular}{llllllll}
\hline \hline type & $m_1,m_2,m_3$ in \textrm{ eV} & 
$\tau_1,\tau_2(\delta=0)$
 & $\tau_1,\tau_2(\delta=0.5)$ &
  $\tau_1,\tau_2(\delta=1.0)$& $\tau_1,\tau_2(\delta=1.5)$& 
$\tau_1,\tau_2(\delta=2.0)$& $\tau_1,\tau_2(\delta=2.5)$ \\
\hline
\hline                                                                 
I & 0.2955, 0.2956, 0.3 & 0.0043,1.0428  & 0.0126, 1.0495 &0.0058, 1.0536  
&-0.0108, 1.0510 &-0.0210, 1.0440 &-0.0153, 1.0394
\\
II & 0.3042, 0.3043, 0.3 & -0.0041,1.0512 & 0.0043,1.0577 &-0.0023, 1.0615 
&-0.0189, 1.0587&-0.0291,1.0518&-0.0235,1.0476
\\
I & 0.0856, 0.0860, 0.1 & 0.0486, 0.9975 & 0.0566, 1.0049 & 0.0489, 
1.0106& 0.0318, 1.0091&0.0219,1.0015&0.0285,0.9953
\\
II & 0.1119, 0.1123, 0.1 & -0.0311, 1.0774 & -0.0226, 1.0835 & -0.0289, 
1.0863&-0.0453, 1.0828&-0.0556,1.0763&-0.0503,1.0731
\\
I & 0.0305, 0.0316, 0.06 & 0.3913, 0.6543 & 0.3873, 0.6748 &0.3578, 
0.7048& 0.3288, 0.7167&0.3258,0.7013&0.3530,0.6720
\\                                                                 
II & 0.0783, 0.0787, 0.06 & -0.0669, 1.1119 & -0.0583, 1.1174 & 
-0.0644,1.1188&-0.0806, 1.1145&-0.0911,1.1085&-0.0860,1.1066
\\
II & 0.0643, 0.0648, 0.04 & -0.1064, 1.1494 & -0.0978, 1.1541 & 
-0.1040, 
1.1538&-0.1203, 1.1483&-0.1307,1.1430&-0.1255,1.1428  \\
II & 0.0541, 0.0548, 0.02 & -0.1747, 1.2115 & -0.1669, 1.2142 &-0.1751, 
1.2095 &-0.1928, 1.2012&-0.2024,1.1976&-0.1951,1.2019 \\
II & 0.0506, 0.0512, 0.005 & -0.2601, 1.2620 & -0.2603, 1.2611 &
 -0.2914, 1.2276&-0.3251, 1.2035&-0.3250,1.2094&-0.2950,1.2369  \\
II & 0.0503, 0.0510, 0.001 & -0.3830, 1.1805 &  & & & & \\
\hline
\hline
\end{tabular}
\end{center}     
\caption[]{Panorama of solutions as $m_3$ is lowered from about the
highest value which is experimentally reasonable to about the
lowest value imposed by the model. For each value of $m_3$,
 predictions are given for $(m_1, m_2)$ and for $(\tau_1, \tau_2)$  
in the cases where 
 $\delta=0, 0.5, 1.0, 1.5, 2.0, 2.5$. All phases are measured in
radians. Here the value $s_{13}^2$
=0.04 was adopted.
 In the type I 
solutions $m_3$ is the largest mass
while in the type II solutions $m_3$ is the smallest mass.} 
\label{trianglesampler}
\end{table}                                                         

    Note that the columns in Table \ref{trianglesampler} 
with $\delta=0$ correspond to the previous case, discussed
in some detail
in section IV of \cite{nsm}. In this case, $m_i^{\prime}$ and
$\tau_i^{\prime}$ respectively coincide with $m_i$ and 
$\tau_i$ so we can identify the vectors of the triangle with
the physical masses and phases. As one decreases
the value of $m_3$ the type I triangle goes from being close to 
equilateral to the degenerate situation with three collinear vectors.
 In this limiting case the vectors representing
neutrino one and neutrino two are approximately equal and add up to 
exactly cancel the vector representing neutrino 3. The precise
orientation of the straight line is due to imposing the constraint
in Eq. (\ref{majoranaphases}). 
 This is actually a CP conserving case 
\cite{realcase}. Then one can find the $m_3$ value
(a little below 0.06 eV) for this situation by
looking for a real solution of $m_1+m_2+m_3=0$ together with   
Eqs. (\ref{massdifferences}) (See Eq. (4.4) of \cite{bfns}).
Clearly there can be no type I solutions below this value of $m_3$.
The type II solutions can exist below this value but similarly 
end (a little below $m_3$ =0.001 eV) when the triangle becomes 
degenerate in a different way. For the type II degenerate triangle,
 the neutrino 1 and neutrino 2 vectors are collinear but oppositely
directed and the small neutrino 3 vector adds to the
 neutrino 1 vector to cancel the neutrino 2 vector. This is also a CP
conserving case. 

    When the effects of $\delta$ not equal to zero are included, it is 
not possible to make a triangle out of the physical neutrino masses and 
phases. The relevant auxiliary triangle is made, as illustrated, using the
primed masses and phases. Thus the limiting values of $m_3$, where
the type I and type II cases each end, correspond to this primed triangle
becoming degenerate.
We can get the limiting value by looking for real solutions
of $\sum_{i}G_im_i$=0, together with Eqs. (\ref{massdifferences}).
 The limiting value, $(m_3)_{min}$ is found to be:
\begin{equation}
(m_3)_{min}^2=\frac{1}{2\alpha}[-\beta - (\beta^2-4\alpha\gamma)^{1/2}],
\label{findm3min}
\end{equation}
where,
\begin{eqnarray}
\alpha &=& (G_1^2+G_2^2-G_3^2)^2-4G_1^2G_2^2,
\nonumber \\
\beta &=& -2(G_1^2+G_2^2-G_3^2)(AG_1^2 \pm B(G_1^2+G_2^2))
+4G_1^2G_2^2(A \pm 2B), \nonumber \\
\gamma &=& (AG_1^2 \pm B(G_1^2+G_2^2))^2 \mp 4G_1^2G_2^2B(A \pm B).
\label{coefficients}
\end{eqnarray}
Here the upper and lower sign choices respectively refer to the type I
and type II cases. 

The computed values of $(m_3)_{min}$  
as a function of $\delta$ are shown in Table \ref{m3minvalues}. 
Looking at this table, one can see why the entries in Table 
\ref{trianglesampler} for $m_3$ =0.001 eV and non-zero values of
$\delta$ are missing. Simply, for those cases, $(m_3)_{min}>$
0.001 eV. Clearly, this correlation of the allowed input values of $m_3$
with the input values of $\delta$ must be respected in studying
the present model. This correlation is imposed by the Ansatz itself.
For large values of $m_3$ there is no constraint from the Ansatz
but Eq. (\ref{cosmobound}) gives an experimental constraint.
It should be remarked that the collinear auxiliary triangles
with non-zero $\delta$ correspond to CP violation. In these
cases the CP violation arises from non trivial phases, $\tau_i$ in 
addition to the assumed non-zero $\delta$. Actually, a better
measure of CP violation involves the (two
independent) phase differences $\tau_i -\tau_j$. These are the
objects essentially related to the internal angles in Fig. \ref{fig:1}.

\begin{table}[htbp]
\begin{center}
\begin{tabular}{lllllll}
\hline \hline type & $(m_3)_{min}$ $(\delta =0)$ in \textrm{ eV} &
 $(m_3)_{min}$ $(\delta =0.5)$
 &  $(m_3)_{min}$ $(\delta =1.0)$ &
  $(m_3)_{min}$ $(\delta =1.5)$&  $(m_3)_{min}$ $(\delta =2.0)$&  
$(m_3)_{min}$ $(\delta =2.5)$         \\
\hline
\hline
I &0.0592716 &0.0590967  & 0.0587178 &0.0584799&0.0586203&0.0589971
\\
II & 0.0006811 & 0.00105461 & 0.0019024 &0.0024636&0.0021294&0.0012723
\\                                                                    
\hline
\hline
\end{tabular}
\end{center}
\caption[]{Minimum allowed value of the input mass, $m_3$ as
a function of the input CP violation phase, $\delta$. These
 correspond to the cases where the triangle in Fig. 1 becomes
degenerate. Here, the choice $s_{13}^2 =$ 0.04 has been made.}
\label{m3minvalues}
\end{table}                                                                

   As noted in ref \cite{nsm}, the possibility of reflecting
the triangle about any line in the plane gives another set of solutions 
 corresponding to reversing the signs of all
the phase differences $\tau_i-\tau_j$. In the present case, where
$\delta$ is not zero, reflecting the ``unphysical" triangle
about any line in the plane will give an alternate solution in which the
$\tau_i^{\prime}-\tau_j^{\prime}$ are reversed in sign.
More specifically, one should 
reverse the signs of the 
first terms on the right hand sides in Eqs. (\ref{thethreetauprimes}). The 
physical phases, $\tau_i$ for this alternate solution
 will then depend on $\delta$ as illustrated in Table 
\ref{flippedtrianglesampler}. 
\begin{table}[htbp]
\begin{center}
\begin{tabular}{llllllll}
\hline \hline type & $m_3$ in \textrm{ eV} &
$\tau_1,\tau_2(\delta=0)$
 & $\tau_1,\tau_2(\delta=0.5)$ &
  $\tau_1,\tau_2(\delta=1.0)$& $\tau_1,\tau_2(\delta=1.5)$&
$\tau_1,\tau_2(\delta=2.0)$& $\tau_1,\tau_2(\delta=2.5)$ \\
\hline
\hline                                                                         
I &  0.3 & -0.0043,-1.0428  &0.0113,-1.0393  &
0.0210, -1.0422 &0.0151, -1.0492 &-0.0015, -1.0536 &-0.0123, -1.0512
\\
II & 0.3 & 0.0041,-1.0512 &0.0196, -1.0475  &0.0291, 
-1.0500
&0.0232, -1.0569 &0.0066, -1.0614 &-0.0040, -1.0593
\\
I & 0.1 & -0.0486, -0.9975 &-0.0326, -0.9947  &-0.0221, 
-0.9992 &-0.0275, -1.0073 &-0.0444, -1.0111 &-0.0560, -1.0070
\\
II & 0.1 & 0.0311, -1.0774 &0.0465, -1.0732 &0.0557, 
-1.0748 &0.0495, -1.0810 &0.0331, -1.0859 &0.0228, -1.0848  
\\
I & 0.06 & -0.3913, -0.6543 &-0.3634, -0.6645  &-0.3310, 
-0.6934 &-0.3246, -0.7149 &-0.3483, -0.7109 &-0.3806,-0.6837
\\
II & 0.06 & 0.0669, -1.1119 &0.0822, -1.1071  &0.0912, 
-1.1073 &0.0849, -1.1127 &0.0685, -1.1180 &0.0584, -1.1183
\\
II & 0.04 & 0.1064, -1.1494 &0.1217, -1.1438  &0.1308, 
-1.1423 &0.1246, -1.1465  &0.1082, -1.1526  &0.0979, -1.1506  
\\   
II & 0.02 & 0.1747, -1.2115 &0.1909, -1.2040  &0.2019, 
-1.1981 &0.1971, -1.1994  &0.1799, -1.2072  &0.1676, -1.2136 \\
II & 0.005 & 0.2601, -1.2620 &0.2853, -1.2447  &
0.3182, -1.2162 &0.3294, -1.2017  &0.3024, -1.2190  &0.2694, -1.2486  \\
II & 0.001 & 0.3830, -1.1805 &  & & & & \\
\hline
\hline
\end{tabular}
\end{center}
\caption[]{Panorama of solutions, using the reflected triangle, 
as $m_3$ is lowered from about the
highest value which is experimentally reasonable to about the 
lowest value imposed by the model. Notice that the predicted masses,
$m_1$ and $m_2$ have not been given since they are the same as in Table 
\ref{trianglesampler}.   } 
\label{flippedtrianglesampler}
\end{table}             

Unlike the $\delta$ = 0 case, the phase differences for the 
reflected triangle solution, are now only approximately the negatives
of those for the original solution. For example, in the case of
a type I triangle with $m_3$ =0.3 and $\delta$ =1.0, 
Table \ref{trianglesampler} shows $\tau_1-\tau_2$=-1.0478 for the original solution while 
Table \ref{flippedtrianglesampler} shows $\tau_1-\tau_2$=+1.0632 for
the reflected triangle solution.

 It should be remarked that the number of decimal places to which
we are calculating is chosen in order to be able to compare various solutions
of the Ansatz with each other for precisely fixed values
of the input mass differences and mixing angles.
The experimental accuracy of the inputs must, of course, be kept
in mind. 

\section{Physical Applications}
     
It is very interesting to note the dependence of our results
on the value of the necessarily small quantity $s_{13}^2$, which
can be seen from the Ansatz Eq. (\ref{paramansatz}) to modulate the 
$\delta$
dependence.
 For this 
purpose let us consider, instead of the value 0.04, the value 0.01.
The resulting analog of Table \ref{trianglesampler} is presented 
in Table \ref{samplerwith0.01}.

\begin{table}[htbp]
\begin{center}
\begin{tabular}{llllllll}
\hline \hline type & $m_3$ in \textrm{ eV} &
$\tau_1,\tau_2(\delta=0)$
 & $\tau_1,\tau_2(\delta=0.5)$ &
  $\tau_1,\tau_2(\delta=1.0)$& $\tau_1,\tau_2(\delta=1.5)$&
$\tau_1,\tau_2(\delta=2.0)$& $\tau_1,\tau_2(\delta=2.5)$ \\
\hline
\hline
I &  0.3 & -0.0043,-1.0428  &0.0063, 1.0445 &0.0046, 1.0455
 &0.0007, 1.0448 &-0.0018, 1.0432  &-0.0006, 1.0420
\\
II & 0.3 & 0.0041,-1.0512 &-0.0020, 1.0528  &-0.0037, 1.0537
&-0.0076, 1.0530 &-0.0101, 1.0514 &-0.0089, 1.0503
\\
I & 0.1 & -0.0486, -0.9975 &0.0505, 0.9994  &0.0486, 1.0008  &0.0446, 
1.0004  &0.0421, 0.9985  &0.0436, 0.9970            
\\
II & 0.1 & 0.0311, -1.0774 &-0.0290, 1.0789  &-0.0306, 1.0796  &-0.0345, 
1.0787  &-0.0370, 1.0772  &-0.0358, 1.0763
\\
I & 0.06 & -0.3913, -0.6543 &0.3900, 0.6597  &0.3816, 0.6681  &0.3738, 
0.6718  &0.3736, 0.6676  &0.3813, 0.6592
\\
II & 0.06 & 0.0669, -1.1119 &-0.0608, 1.1132  &-0.0634, 1.1136  &-0.0702, 
1.1125  &-0.0727, 1.1111   &-0.0716, 1.1106
\\
II & 0.04 & 0.1064, -1.1494 &-0.1043, 1.1505  &-0.1058, 1.1504 &-0.1097, 
1.1492  &-0.1122, 1.1479  &-0.1111, 1.1478
\\
II & 0.02 & 0.1747, -1.2115 &-0.1728, 1.2122   &-0.1748, 1.2110  &-0.1789, 
1.2091  &-0.1813, 1.2082  &-0.1797, 1.2091  
\\   
II & 0.005 & 0.2601, -1.2620 &-0.2605, 1.2602  &-0.2673, 1.2538
 &-0.2744, 1.2486  &-0.2750, 1.2496  &-0.2687, 1.2558  \\
II & 0.001 & 0.3830, -1.1805 &0.4036, 1.1593  &-0.4828, 1.0840 & & 
&-0.4252, 1.1420 \\
\hline
\hline
\end{tabular}
\end{center}
\caption[]{Panorama of solutions as in Table \ref{trianglesampler}
but with $s_{13}^2$ =0.01.
 Notice that the predicted masses,
$m_1$ and $m_2$ have not been given since they are the same as in Table   
\ref{trianglesampler}.   }
\label{samplerwith0.01}
\end{table}

Notice that Table \ref{samplerwith0.01} has fewer missing solutions for 
the case
$m_3$ =0.001 than does Table \ref{trianglesampler}. This is 
because  decreasing $s_{13}^2$ brings the $G_i$ in Eq. 
(\ref{definemprime})
closer to unity, which in turn brings the physical neutrino masses
closer to the auxiliary $m_i^{\prime}$. 
 The modified lower limits for
$m_3$ are illustrated in Table \ref{m3minvalueswith0.01}.

\begin{table}[htbp]
\begin{center}
\begin{tabular}{lllllll}
\hline \hline type & $(m_3)_{min}$ $(\delta =0)$ in \textrm{ eV} &
 $(m_3)_{min}$ $(\delta =0.5)$
 &  $(m_3)_{min}$ $(\delta =1.0)$ &
  $(m_3)_{min}$ $(\delta =1.5)$&  $(m_3)_{min}$ $(\delta =2.0)$&
$(m_3)_{min}$ $(\delta =2.5)$         \\
\hline
\hline
I &0.0592716 &0.0592263  &0.0591312  &0.0590735  &0.0591074  &0.0592009
\\
II & 0.0006811 &0.0007755  &0.0009761  &0.0010993  &0.0010268  &0.0008287
\\
\hline
\hline
\end{tabular}
\end{center}          
\caption[]{Minimum allowed value of the input mass, $m_3$ as
a function of the input CP violation phase, $\delta$
as in Table \ref{m3minvalues} but with $s_{13}^2$ =0.01.} 
\label{m3minvalueswith0.01}
\end{table}

    The implications of this model are relevant to experiments 
which are designed to search for evidence of neutrinoless double beta 
decay. The amplitudes for these processes contain a factor, $m_{ee}$,
which is independent
of the nuclear wave functions. Its magnitude is given by,
\begin{equation}
|m_{ee}| = |m_1(K_{exp11})^2e^{-2i\tau_1}+m_2(K_{exp12})^2e^{-2i\tau_2}
+m_3(K_{exp13})^2e^{-2i\tau_3}|,
\label{ndbd}
\end{equation}                                                              
which appears to require, for its evaluation, a full knowledge of the 
neutrino masses, mixing angles and CP violation phases. The
present experimental bound \cite{ndbdexpt} on this quantity is
\begin{equation}
|m_{ee}| < (0.35 \rightarrow 1.30) \textrm{ eV}, \label{ndbd1}
\end{equation}
A very recent review of neutrinoless double beta decay is
given in ref. \cite{aetal}.                                       
Using the general parameterization of Eq. (\ref{fullmixingmatrix}) one 
finds.
\begin{equation}
m_{ee}=\sqrt{~C^{2}+D^{2}},
\label{mee1}
\end{equation}
wherein,
\begin{eqnarray}
C&=&m_{1}\left( c_{12}c_{13}\right) ^{2}+m_{2}\left( s_{12}c_{13}\right)
^{2}\cos \left[ 2\left( \tau _{2}-\tau _{1}\right) \right] +m_{3}\left(
s_{13}\right) ^{2}\cos \left[ 2\left( \tau _{3}-\tau _{1}+\delta \right)
\right], \nonumber \\
D&=&m_{2}\left( s_{12}c_{13}\right) ^{2}\sin \left[ 2\left( \tau _{2}-\tau
_{1}\right) \right] +m_{3}\left( s_{13}\right) ^{2}\sin \left[ 2\left(
\tau
_{3}-\tau _{1}+\delta \right) \right].
\label{meecoeffs}
\end{eqnarray}
The dependence of $|m_{ee}|$ on the input values of $m_3$ and $\delta$,
obtained by using the Ansatz of present interest, 
is displayed in Table. \ref{mee} for the same choices as in Table
\ref{trianglesampler}. There is noticeable dependence on the
input CP phase $\delta$ for the larger values of $m_3$.
\begin{table}[htbp]
\begin{center}
\begin{tabular}{llllllll}
\hline \hline type & $m_1,m_2,m_3$ in \textrm{ eV} &
$|m_{ee}|(\delta=0)$
 & $|m_{ee}|(\delta=0.5)$ &
  $|m_{ee}|(\delta=1.0)$& $|m_{ee}|(\delta=1.5)$&
$|m_{ee}|(\delta=2.0)$& $|m_{ee}|(\delta=2.5)$ \\
\hline
\hline                                                            
I & 0.2955, 0.2956, 0.3 & 0.164 & 0.174 & 0.183 & 0.181&0.170&0.162 \\
II & 0.3042, 0.3043, 0.3 & 0.167 & 0.177 & 0.185 & 0.183&0.172&0.164 \\
I&0.0856, 0.0860, 0.1&0.051&0.055&0.057&0.057&0.054&0.051 \\
II&0.1119, 0.1123, 0.1&0.058&0.062&0.065&0.064&0.060&0.057 \\
I & 0.0305, 0.0316, 0.06 & 0.026 & 0.028 & 0.029 & 0.030&0.029&0.027 \\
II & 0.0783, 0.0787, 0.06  & 0.038 & 0.040 & 0.042 & 0.041&0.039&0.037 \\
II&0.0643, 0.0648, 0.04&0.029&0.031&0.032&0.031&0.029&0.028 \\
II&0.0541, 0.0548, 0.02&0.022&0.023&0.023&0.023&0.022&0.021 \\
II & 0.0506, 0.0512, 0.005 & 0.019 & 0.020 & 0.020 & 0.019 &0.019&0.019\\
II  & 0.0503, 0.0510, 0.001 &0.019&  &  &  & & \\
\hline
\hline
\end{tabular}
\end{center}
\caption[]{The neutrinoless double beta decay amplitude
 factor, $|m_{ee}|$
in eV as a function of the input CP violation phase, $\delta$. 
 Here, the choice $s_{13}^2 =$ 0.04 has been made.}
\label{mee}   
\end{table}

     For the reflected triangle 
solutions discussed above, the predictions of $|m_{ee}|$ are given below
in Table \ref{meeflipped}. Again there is a noticeable dependence
on $\delta$ for the larger values of $m_3$. However, the peak values occur
at different values of $\delta$ compared to Table VI.

\begin{table}[htbp]
\begin{center}
\begin{tabular}{llllllll}
\hline \hline type & $m_3$ in \textrm{ eV} &
$|m_{ee}|(\delta=0)$
 & $|m_{ee}|(\delta=0.5)$ &
  $|m_{ee}|(\delta=1.0)$& $|m_{ee}|(\delta=1.5)$&
$|m_{ee}|(\delta=2.0)$& $|m_{ee}|(\delta=2.5)$ \\
\hline
\hline
I & 0.3 & 0.164 &0.161 &0.167  &0.178 &0.183 &0.177 \\
II & 0.3 & 0.167 &0.163  &0.169  &0.180 &0.186 &0.180 \\
I& 0.1&0.051&0.051 &0.053 &0.056 &0.058  &0.056 \\
II& 0.1&0.058&0.057 &0.059  &0.063  &0.065  &0.063 \\
I &  0.06 & 0.026 &0.027  &0.029  &0.030 &0.030 &0.028 \\
II &  0.06  & 0.038 &0.037  &0.038  &0.040  &0.042 &0.041 
\\
II& 0.04&0.029&0.028 &0.029 &0.030 &0.032 &0.031 \\
II& 0.02&0.022&0.021 &0.022  &0.022 &0.023 &0.023 \\
II & 0.005 & 0.019 &0.019  &0.019  &0.019  &0.019 &0.020 
\\
II  &  0.001 &0.019&  &  &  & & \\
\hline
\hline
\end{tabular}
\end{center}  
\caption[]{The neutrinoless double beta decay amplitude
 factor $|m_{ee}|$
in eV as a function of the input CP violation phase, $\delta$,
using the reflected triangle.
 Notice that the predicted masses,
$m_1$ and $m_2$ have not been given since they are the same as in Table
\ref{mee}.   }                                   
\label{meeflipped}
\end{table}

 The effects of lowering $s_{13}^2$ to 0.01 are finally illustrated below,
for the non-reflected triangle case, in Table \ref{meewith0.01}.

\begin{table}[htbp]
\begin{center}
\begin{tabular}{llllllll}
\hline \hline type & $m_3$ in \textrm{ eV} &
$|m_{ee}|(\delta=0)$
 & $|m_{ee}|(\delta=0.5)$ &
  $|m_{ee}|(\delta=1.0)$& $|m_{ee}|(\delta=1.5)$&
$|m_{ee}|(\delta=2.0)$& $|m_{ee}|(\delta=2.5)$ \\
\hline
\hline
I & 0.3 &0.177  &0.180  &0.181  &0.182  &0.179  &0.176  \\
II & 0.3 &0.179  &0.182  &0.184  &0.183   &0.181  &0.179  \\
I& 0.1&0.056 &0.057 &0.057 &0.057 &0.056  &0.056 \\ 
II& 0.1&0.063  &0.064  &0.064  &0.064  &0.063  &0.062 \\
I &  0.06 &0.029  &0.029  &0.030  &0.030 &0.030 &0.029 \\
II &  0.06  &0.041  &0.041  &0.042  &0.041  &0.041 &0.040
\\
II& 0.04&0.031  &0.031  &0.031  &0.031  &0.031 &0.031 \\
II& 0.02&0.023  &0.023  &0.023  &0.023   &0.023  &0.023 \\
II & 0.005 &0.020  &0.020  &0.020  &0.020  &0.020  &0.020
\\
II  &  0.001 &0.020  &0.020  &0.020  &  & &0.020 \\
\hline               
\hline
\end{tabular}
\end{center}
\caption[]{The neutrinoless double beta decay amplitude
 factor $|m_{ee}|$
in eV as a function of the input CP violation phase, $\delta$,
as in Table \ref{mee} but with $s_{13}^2$ = 0.01.
 Notice that the predicted masses,
$m_1$ and $m_2$ have not been given since they are the same as in Table
\ref{mee}.   }     
\label{meewith0.01}
\end{table} 

    The main conclusion of this model for neutrinoless 
double beta decay, obtained by looking at all three tables
above and noting the smooth dependence of $|m_{ee}|$ on
the inputs $m_3$ and $\delta$ for each of the type I and type II
solutions,
 is that $|m_{ee}|$ should satisfy the
restrictive approximate bounds:
\begin{equation}
     0.020 eV < |m_{ee}| < 0.185  eV .
\label{meebounds}
\end{equation}
Here, the lower bound is intrinsic to the model but the upper bound  
reflects the experimental bound on the sum of neutrino masses
quoted in Eq. (\ref{cosmobound}) and might be improved upon.
We also note that, when both type I and type II solutions exist
for a given value of $m_3$, the type II solution gives somewhat larger
$|m_{ee}|$. The main dependence of $|m_{ee}|$ is, of course, on
 the input parameter $m_3$.    

   We plan to discuss elsewhere other physical applications
including beta decay end point spectra \cite{fps} and leptogenesis
\cite{fy} using the approach of ref. \cite{nsm}.

\section{ symmetrical parameterization}

In the parameterization of the leptonic mixing matrix given by
Eq.(\ref{fullmixingmatrix}), which is similar to the one usually adopted, 
there
appears to be an important distinction between the phase $\delta$
and the phases $\tau_i$ in the sense that only $\delta$ survives when
one considers the ordinary (overall lepton number conserving)
neutrino oscillation experiments. However, this distinction may be 
preserved in a different way while using a more symmetrical
parameterization. Thus consider,  
\begin{equation}
K_S=\omega_{23}(\theta_{23},\phi_{23})\omega_{13}(\theta_{13},\phi_{13})
\omega_{12}(\theta_{12},\phi_{12}),
\label{sympar}
\end{equation}
which contains the same mixing angles, $\theta_{ij}$ as before but now has 
the three associated CP violation phases, $\phi_{ij}$. The $\omega_{ij}
(\theta_{ij},\phi_{ij})$ were defined as in Eq. (\ref{onetwotransf}). 
Writing out the 
whole matrix yields,  
\begin{equation}
K_S=\left[ \begin{array}{c c c}
c_{12}c_{13}&s_{12}c_{13}e^{i{\phi_{12}}}&s_{13}e^{i{\phi_{13}}}\\
-s_{12}c_{23}e^{-i{\phi_{12}}}-c_{12}s_{13}s_{23}e^{i({\phi_{23}}-{\phi_{13}})}
&c_{12}c_{23}-s_{12}s_{13}s_{23}
e^{i({\phi_{12}}+{\phi_{23}}-{\phi_{13}})}&c_{13}s_{23}e^{i{\phi_{23}}}\\
s_{12}s_{23}e^{-i({\phi_{12}}+{\phi_{23}})}-c_{12}s_{13}c_{23}e^{-i{\phi_{13}}}
&-c_{12}s_{23}e^{-i{\phi_{23}}}-
s_{12}s_{13}c_{23}e^{i({\phi_{12}}-{\phi_{13}})}&c_{13}c_{23}\\
\end{array} \right].
\label{writeout}
\end{equation}

To relate this form to the previous one, we can use the identity
\cite{sv},
\begin{equation}
\omega^{-1}_0(\tau)K_S\omega_0(\tau)=
\omega_{23}(\theta_{23}, -\tau_2+\phi_{23}+\tau_3)
\omega_{13}(\theta_{13}, -\tau_1+\phi_{13}+\tau_3)     
\omega_{12}(\theta_{12}, -\tau_1+\phi_{12}+\tau_2),
\label{transformKS}     
\end{equation}
where the diagonal matrix of phases, $\omega_0(\tau)$ was defined
in Eq. (\ref{majoranaphases}).
Now choose the $\tau_i$'s (two are independent) so that the transformed
(12) and (23) phases vanish. Then, with the identifications,
\begin{eqnarray}
\phi_{12}&=&\tau_1-\tau_2,
\nonumber \\
\phi_{23}&=&\tau_2-\tau_3,
\nonumber \\
\phi_{13}&=&\tau_1-\tau_3 -\delta,
\label{relatingphases}    
\end{eqnarray}
we notice that $K_S$ is related to $K$ of
 Eq. (\ref{fullmixingmatrix}) as 
\begin{equation}
K_S=\omega_0(\tau)K.
\label{relateK&KS}
\end{equation}
Since $K$ sits in the weak interaction
Lagrangian, Eq. (\ref{weakinteraction}) with the charged lepton field
row vector on its left, all physical results will be unchanged
if $K$ is multiplied by a diagonal matrix of phases on its left.
Thus $K$ and $K_S$ are equivalent and the relations between the CP
violation phases of the two parameterizatons are given by
Eq. (\ref{relatingphases}). By construction, $\delta$ is the  
only CP violation phase which can appear in the description of ordinary
 neutrino oscillations. Solving the Eqs. (\ref{relatingphases}) for
$\delta$ in terms of the three $\phi_{ij}$'s gives,
\begin{equation}
\delta =\phi_{12} +\phi_{23}-\phi_{13},
\label{invariantphase}
\end{equation}
which shows that in the symmetrical parameterization, the
``invariant phase"  \cite{gs} combination $
I_{123} =\phi_{12} +\phi_{23}-\phi_{13}$
is the object which measures CP violation for ordinary neutrino
oscillations. It has the desired property of intrinsically
spanning three generations, as needed for CP violation in ordinary
neutrino oscillation or in the quark mixing analog. Furthermore,
it can be seen \cite{gjs} to have an interesting mathematical
structure and to be
 useful for extension to the case where there are more than three
 generations of fermions

    The convenience of the symmetrical parameterization can already be 
seen in Eqs. (\ref{meecoeffs}) needed for obtaining $|m_{ee}|$. These
equations simplify when one observes that the combinations of
phases occurring within them are simply $\phi_{12}$ and $\phi_{13}$.
These are evidently the two phases which describe the coupling
of the first lepton generation.   
Similarly, the estimates of CP violation needed for the treatment
of leptogenesis made in section V of \cite{nsm} also simplify
when expressed in terms of the $\phi_{ij}$.

\section{Summary and discussion}

 Assuming that the squared
neutrino mass differences and the three (CP conserving)
lepton mixing angles are known, 
the mass of one neutrino and three CP violation phases
remain to be determined.
The complementary Ansatz, $Tr(M_\nu)$ =0, provides two
real conditions on the parameters of the lepton system. 
Here, we took $m_3$ and $\delta$ as input parameters
and then determined the other two masses and the other two
CP violation phases according to this Ansatz. A 
geometric algorithm was presented based on
a modification of an earlier treatment \cite{nsm}
in which the Dirac CP violation phase $\delta$
was neglected, but the other two Majorana 
type CP violation phases were retained. That is a reasonable 
first approximation because
the effect of delta is always suppressed by $s_{13}$
which is known experimentally to be small. However, there
is great interest in the determination of $\delta$ 
so it should not be ignored. Additionally
in ref. \cite{nsm} it was suggested that small CP
violation scenarios might be close to the physical case.
The present algorithm is exact and does not require the
 assumption that any parameters are small.

     The Ansatz yields a characteristic pattern for
 the neutrino masses and the CP violation phases. Because
 $s_{13}^2$ is small, the main cause
of change is the assumed value of the input parameter $m_3$.
First one notes that the small experimental value of
A in Eq. (\ref{massdifferences}) always forces the neutrino 1 and neutrino 
2 masses
to be almost degenerate (See table \ref{trianglesampler}).
For the largest allowed (from the cosmological bound
Eq. (\ref{cosmobound})) value of $m_3$, around 0.3 eV, there is an
approximate three-fold degeneracy of all the neutrino masses.
This is understandable since when the mass scale becomes large,
both A and B can be considered negligible. Then
the triangle of Fig. 1 becomes approximately equilateral. The
internal angles of the triangle approximately measure the strength
of the CP phases $\tau_i-\tau_j$ and are clearly large in this situation.
As $m_3$ decreases, a point around 0.06 eV is reached at which
 the type I solutions 
($m_3$ largest) no longer exist. At this point the CP violation vanishes
for the $\delta$ =0 case and becomes small when $\delta \ne$0. Also at this
point the almost degenerate neutrino 1 and neutrino 2 masses decrease to about
half the neutrino 3 mass in the type I case. In the present model
 neutrinos 1 and
 2 never go below about half the mass of neutrino 3. The situation is a little
different for the type II cases ($m_3$ smallest). The type II solutions
exist from a maximum of $m_3$ about 0.3 eV to a minimum of about 0.001 eV.
At the minimum the CP violation ceases for $\delta$ = 0 and has small effects
when $\delta \ne$ 0. Furthermore, at the minimum neutrinos 1 and 2 are about
50 times heavier than neutrino 3. Thus a possible hierarchy of 
neutrino masses can only exist in one way for the present model,
with $m_3$ considerably smaller than the other two.
 
   Some further technical details of the model were displayed in Tables
II -IV. These describe the $\delta$ dependence of the limiting values of
the input parameter $m_3$ just mentioned, the solution
 corrresponding to a reflected triangle
and the effect of varying $s_{13}^2$.

    The model might be handy for getting an idea about the range of predictions
for various leptonic phenomena, since it gives a plausible two parameter
complete set of
neutrino masses, mixing angles and CP violation phases. For example,
 in section V an application is made to the quantity $|m_{ee}|$
which characterizes neutrinoless double beta decay. The results
for about the largest allowed $s_{13}^2$, the reflected solution
 and a smaller $s_{13}^2$ choice are shown in Tables VI, VII
and VIII. It may be noted that the solutions vary rather smoothly
with $m_3$ and $\delta$ for a solution of given type. Thus,
even though one might initially expect the result of allowing a two 
parameter 
family choice to be rather weak, it turns out that one gets fairly 
restrictive upper and lower bounds on $|m_{ee}|$ as expressed
in  Eq. (\ref{meebounds}). These approximate bounds may be considered 
also as a test of the present model. Of course, any
direct determination of a
neutrino mass, say from a beta decay endpoint experiment, will also
provide a test of the model.

    In section V we took up a question which is independent
of the present Ansatz. How should one parameterize the lepton mixing
matrix?  Of course, this is fundamentally a question of choice.
However, we notice in the present work that the physical
 quantities we calculate,
depend in the simplest way not on the conventional phases $\delta$ 
and $\tau_i$ (where $\sum \tau_i =0$) but on the quantities
$\phi_{ij}$ given in Eq. (\ref{relatingphases}). Such a dependence
arises naturally if one uses the alternative mixing matrix parameterization given
by $K_S$ in Eq. (\ref{sympar}).
 This may be understood
physically in the following way. The $\phi_{ij}$'s by definition (see
Eq.(\ref{onetwotransf}) for example) span two generations.
 It is known \cite{svandgkm} that CP violation begins at the two
generation level for Majorana neutrinos. Thus it seems
 appropriate that the $\phi_{ij}$'s should appear. However
 if $\delta$ = 0, the three $\phi_{ij}$'s are not linearly independent
according to Eq. (\ref{invariantphase}).
This takes care of the two Majorana phases. When $\delta \ne 0$ the
three $\phi_{ij}$'s
are of course independent and the ``invariant phase" combination $I_{123}$
discussed some time ago \cite{gs,gjs} intrinsically spans three
generations, as expected for "Dirac" type CP violation.

\section*{Acknowledgments}
\vskip -.5cm
  S.S.M would like to thank the Physics Department at Syracuse University
 for their hospitality.
 The work of S.N is supported by National Science foundation grant No. 
PHY-0099.
The work of J.S. is supported in part by the U. S. DOE under
Contract no. DE-FG-02-85ER 40231.


\begin{thebibliography}{9}

\bibitem{kamland}KamLAND collaboration, K. Eguchi et al, Phys. Rev.
Lett. {\bf 90}, 021802 (2003).

\bibitem{sno}SNO collaboration, Q. R. Ahmad et al, arXiv:nucl-ex/
0309004.

\bibitem{k2k}K2K collaboration, M. H. Ahn et al, Phys. Rev. Lett. {\bf
90}, 041801 (2003).

\bibitem{nos}For recent reviews see, for examples,
S. Pakvasa and J. W. F. Valle, special issue on neutrino, Proc.
Indian Natl. Sci. Acad. Part A {\bf 70}, 189 (2004); 
arXiv:hep-ph/0301061 and
V. Barger, D. Marfatia and K. Whisnant, Int. J. Mod. 
Phys. E {\bf 12}, 569 (2003); arXiv:hep-ph/0308123.   
                                           
\bibitem{lsnd}LSND collaboration, C. Athanassopoulos et al, Phys. Rev.
Lett. {\bf 81}, 1774 (1998).

\bibitem{mbc}Recent summaries of the status of this experiment are
given by H. Ray, arXiv:hep-ex/0411022 and A. Aguilar-Arevalo,
arXiv:hep-ex/0408074. See also R. Tayloe [MiniBooNE Collaboration],
Nucl. Phys. Proc. Suppl. {\bf 118}, 157 (2003).

\bibitem{sv}J. Schechter and J. W. F. Valle, Phys. Rev. D {\bf 22}, 2227
(1980).
 
\bibitem{bhp}S. M. Bilenky, J. Hosek and S. T. Petcov, Phys. Lett. {\bf 94B}, 495
(1980).

\bibitem{dknot}M. Doi, T. Kotani, H. Nishiura, K. Okuda and E. Takasugi, Phys. Lett.
{\bf 102B}, 323 (1981).

\bibitem{svandgkm}J. Schechter and J. W. F. Valle, Phys. Rev. {\bf D23}, 1666
(1981); A. de Gouvea, B. Kayser and R. N. Mohapatra, Phys. Rev.
{\bf D67}, 053004 (2002).

\bibitem{mstv} M. Maltoni, T. Schwetz, M. A. Tortola and J. W. F. Valle,
Phys. Rev. D {\bf 68}, 113010 (2003). For the choice $m_2>m_1$ see
A. de Gouvea, A. Friedland and H. Murayama, Phys. Lett.
{\bf B490}, 125 (2000).           


\bibitem{bfns}D. Black, A. H. Fariborz, S. Nasri and J. Schechter,  
Phys. Rev. {\bf D62}, 073015 (2000).

\bibitem{hz}X.-G. He and A. Zee, Phys. Rev. {\bf D68}, 037302, (2003).

\bibitem{ro}W. Rodejohann, Phys. Lett. {\bf B579}, 127 (2004).

\bibitem{nsm}S. Nasri, J. Schechter and S. Moussa, Phys. Rev. {\bf D70},
053005 (2004).

\bibitem{nsm2}S. Nasri, J. Schechter and S. Moussa, arXiv: 
hep-ph/0406243, to be published in the Proceedings of MRST 2004,
Concordia University, Montreal, May 2004.

\bibitem{metal}R. N. Mohapatra et al, arXiv:hep-ph/0412099.

\bibitem{cosmobound}D. N. Spergel et al, Astrophys. J. Suppl. 
{\bf 148}: 175 (2003);
S. Hannestad, JCAP {\bf 0305}: 004 (2003).                              

\bibitem{realcase}See the discussion in section IV of \cite{bfns}
above and L. Wolfenstein, Phys. Lett. {\bf 107B}, 77 (1981). 

\bibitem{ndbdexpt}H. V. Klapdor-Kleingrothaus {\it et al},
Eur. Phys. J. A {\bf 12}, 147 (2001).

\bibitem{aetal}C. Aalseth et al, arXiv:hep-ph/0412300.

\bibitem{fps}Y. Farzan, O. L. G. Peres and A. Yu. Smirnov,
Nucl. Phys. B {\bf 612} (2001).

\bibitem{fy}M. Fukugita and T. Yanagida, Phys. Lett.
{\bf B174}, 45 (1986). The original                      
baryogenesis mechanism is given in A. D. Sakharov,
Pis'ma Zh. Eksp. Teor. Fiz. {\bf 5}, 24 (1967) (JETP Lett. {\bf 5}, 24 
(1967)).
                                      

\bibitem{gs}M. Gronau and J. Schechter, Phys. Rev. Lett. {\bf 54},
385 (1985); erratum {\bf 54}, 1209 (1985).

\bibitem{gjs}M. Gronau, R. Johnson and J. Schechter, Phys. Rev.
{\bf D32}, 3062 (1985). The sense in which $I_{123}$
is entitled to be called an invariant phase for the quark mixing matrix
or equivalently for the ordinary neutrino oscillations is discussed in
Eqs. ($2^{\prime}$), ($4^{\prime}$), (13a) and (13b) of this reference.



\end{thebibliography}
\end{document}